# Emergence of a Non-van der Waals Magnetic Phase in a van der Waals Ferromagnet

Bikash Das[1,#], Subrata Ghosh[1,#], Shamashis Sengupta[2], Pascale Auban-Senzier[3], Miguel Monteverde[3], Tamal Kumar Dalui[1], Tanima Kundu[1], Rafikul Ali Saha[4], Sujan Maity[1], Rahul Paramanik[1], Anudeepa Ghosh[1], Mainak Palit[1], Jayanta K Bhattacharjee[1], Rajib Mondal[5], and Subhadeep Datta[1,*]

[1]*School of Physical Sciences, Indian Association for the Cultivation of Science, 2A & B Raja S. C. Mullick Road, Jadavpur, Kolkata - 700032, India*

[2]*Université Paris-Saclay, CNRS/IN2P3, IJCLab, 91405 Orsay, France*

[3]*Université Paris-Saclay, CNRS, Laboratoire de Physique des Solides, 91405, Orsay, France*

[4]*Department of Microbial and Molecular Systems, Centre for Membrane Separations, Adsorption, Catalysis and Spectroscopy for Sustainable Solutions (cMACS), KU Leuven, Celestijnenlaan 200F, 3001 Leuven, Belgium*

[5]*UGC-DAE Consortium for Scientific Research, Kolkata Centre, Bidhannagar, Kolkata 700 106, India*

\# Authors equally contributed to this work
*Email: sspsdd@iacs.res.in

**Abstract**

**Manipulation of long-range order in two-dimensional (2D) van der Waals (vdW) magnetic materials (*e.g.,* $CrI_3$, $CrSiTe_3$ etc.), exfoliated in few-atomic layer, can be achieved *via* application of electric field, mechanical-constraint, interface engineering, or even by chemical substitution/doping. Usually, active surface oxidation due to the exposure in the ambient condition and hydrolysis in the presence of water/moisture causes degradation in magnetic nanosheets which, in turn, affects the nanoelectronic**

/spintronic device performance. Counterintuitively, our current study reveals that exposure to the air at ambient atmosphere results in advent of a stable nonlayered secondary ferromagnetic phase in the form of $Cr_2Te_3$ ($T_{C2}$ ~ 160 K) in the parent vdW magnetic semiconductor $Cr_2Ge_2Te_6$ ($T_{C1}$ ~ 69 K). In addition, the magnetic anisotropy energy (MAE) enhances in the hybrid by an order from the weakly anisotropic pristine $Cr_2Ge_2Te_6$ crystal, increasing the stability of the FM ground state with time. Comparing with the freshly prepared $Cr_2Ge_2Te_6$, the coexistence of the two ferromagnetic phases in the time elapsed bulk crystal is confirmed through systematic investigation of crystal structure along with detailed dc/ac magnetic susceptibility, specific heat, and magneto-transport measurement. To capture the concurrence of the two ferromagnetic phases in a single material, Ginzburg-Landau theory with two independent order parameters (as magnetization) with a coupling term can be introduced. In contrast to rather common poor environmental stability of the vdW magnets, our results open possibilities of finding air-stable novel materials having multiple magnetic phases.

**Keywords:** 2D vdW materials, ferromagnetic ordering, magnetic anisotropy, magneto-transport

## I. INTRODUCTION

External perturbation in the form of electric field, hydrostatic pressure, light beams or chemical doping can manipulate the ground state of a two-dimensional (2D) magnet towards realization of multiphase spintronic logic.[1] For the survival of long-range 2D magnetic ordering at finite temperature, which is restricted due to thermal fluctuations in a 2D isotropic ferromagnet,[2] magnetocrystalline anisotropy (~ 1 meV/magnetic atom) is a critical factor through interlayer coupling.[3] In this context, 2D semiconducting van der Waals (vdW) ferromagnets such as $Cr_2Si_2Te_6$,[4] $Cr_2Ge_2Te_6$,[5] and $CrI_3$[6] are favourable to be applicable in the field of spintronic and data storage devices due to their long-range ferromagnetic ordering which persists in atomically thin flakes. Experimentally, magnetism in exfoliated layers can be probed by different techniques such as magneto-optical Kerr effect (MOKE),[5,7] reflective magnetic circular dichroism (RMCD),[8] magnetic force microscopy (MFM)[9] etc. However, air-stability (*e.g.*, degradation of $CrI_3$ flakes in the air in 15 min) and aging effects (change of coercive field with time) due to hydrolysis, surface oxidation or light irradiation are the significant obstacles for device applications, even for the fundamental research.

In the family of ferromagnetic $CrXTe_3$ (X = Si, Ge), $Cr_2Ge_2Te_6$ has emerged as a promising material due to its ability to exhibit intrinsic characteristics of both semiconductor (band gap 0.7 eV) and long-range ferromagnetic order ($T_C$ ~ 67 K) with better chemical/air stability to resist environmental effects on the surface degradation (by $TeO_x$ formation) than the other members.[5,10–13] In $Cr_2Ge_2Te_6$, individual atomic layers are stacked together along the crystallographic *c* axis via vdW interaction. It is observed that vdW gap reduces and the cleavage energy enhances with increasing the radius of X atom. According to previous theoretical predictions,[14,15] Cr-Te-Cr super-exchange interaction determines the ferromagnetic (FM) ground state of these materials whereas, Cr-Cr direct exchange interaction favours antiferromagnetism (AFM) which is inversely proportional to the Cr-Cr distance.[16] Note that

$Cr^{3+}$ ions with spin S = 3/2 which are octahedrally coordinated by Te ligands form a 2D honeycomb like magnetic lattice in the *ab* plane and the unit cell is generated by three $Cr_2Ge_2Te_6$ layers stacked in ABC sequence along the *c* axis *via* vdW interaction.[17] Moreover, Xing et al.[7] reported that FM is maintained upon exfoliating the bulk $Cr_2Ge_2Te_6$ crystals down to few-atomic layer which also shows an excellent tunability with electrostatic gate in the three-terminal device geometry. Nevertheless, the type of magnetic interaction of $Cr_2Ge_2Te_6$ is still controversial: based on spin-wave theory, $Cr_2Ge_2Te_6$ is proposed to be a nearly ideal 2D Heisenberg ferromagnet,[5,18] the critical behavior of $Cr_2Ge_2Te_6$ was observed to follow the tricritical mean-field model[19] whereas the critical exponents deduced by Liu *et al.*[12] and Liu *et al.*[10] are consistent with a 2D-Ising model. Interestingly, a recent study shows $Cr_2Ge_2Te_6$ nanosheets can be prepared from non-vdW ferromagnetic $Cr_2Te_3$ template ($T_C$ ~ 180 K) by cation exchange (substitution of $Cr^{3+}$ with $Ge^{4+}$) with a possibility of the coexistence of the two phases.[20] Even, electrochemical intercalation of organic molecules can significantly increase the $T_C$ with altering the weak superexchange interaction in the pristine $Cr_2Ge_2Te_6$ to strong double-exchange interaction in the hybrid.[21] That being the case, presence of multicritical points resulting in competition between magnetic exchange interactions (symmetric or antisymmetric), as reported in quasi-two-dimensional (Q2D) bilayer manganite or multiferroic $Cu_2OSeO_3$, is elusive for future magnetic memory devices.[22,23] Even, systems with interacting magnetic units, like 2D ferrimagnetic ludwigite as frustrated magnet, can be useful for realizing collective quantum modes.[24] Although not a conventional approach, aging may also promote novel spin glass phase coexisting with antiferromagnetism in topologically frustrated pyrochlore.[25] However, for the application, it is highly desirable to search for multiple magnetic phases to exist in a single material with high $T_c$ and importantly, scalability as a channel material in electronic/spintronic device geometry. Foreseeably, vdW few-layer magnets with concurrent stable multiple magnetic phases in ambient condition can be utilized as an active material for tunable vdW spin memory.

In this work, we have systematically investigated the structural, electronic and magnetic properties of air-exposed vdW $Cr_2Ge_2Te_6$ bulk crystals. Pristine single crystals of $Cr_2Ge_2Te_6$ ($T_C \sim 67$ K) were intentionally kept in the ambient atmospheric condition for a long-time duration of about 12 months. In the time-elapsed crystal, in addition to the pristine phase, a secondary ferromagnetic phase with $T_C \sim 160$ K can be observed in the temperature dependent dc magnetization data and can be identified with non-vdW FM $Cr_2Te_3$. Moreover, the air-exposed sample shows stronger magnetic anisotropy than the pristine sample and sustains it 2D nature with time. The study of ac magnetic susceptibility under different frequencies excludes the possibility of spin glass state in the modified sample. Low-temperature anisotropic magnetoresistance shows clear signature of departure from the pristine $Cr_2Ge_2Te_6$ and can be deconvoluted with effects originating from two different magnetic anisotropies in a hybrid material consisting of $Cr_2Ge_2Te_6$ and $Cr_2Te_3$. Analytically, concurrence of the two ferromagnetic phases in a single material can be realized by adopting Ginzburg-Landau theory with two independent order parameters (as magnetization) with a coupling term. Contrary to the common perception, our results with aged, but not decrepit samples open possibilities of multistate logic consisting of emergent magnetic phases with unaffected ferromagnetism in the parent material keeping its layered nature intact.

## II. EXPERIMENTAL

Single crystals of $Cr_2Ge_2Te_6$ sample were grown by the self-flux technique[16] using pure constituent elements (purity > 99.99 %). Cr, Ge, and Te powders were mixed with a molar ratio of 1: 2: 6 and subsequently, the mixture was sealed in an evacuated quartz tube. The quartz tube was heated to 1373 K for 3 hours in a box furnace followed by slow cooling to 973 K at a rate of 1 K/min. The extra flux was removed by centrifuge technique at high temperature.

These single crystals were kept in the ambient atmosphere for about 12 months and subsequently, magnetic and structural properties of the air-exposed $Cr_2Ge_2Te_6$ crystals were examined by several experimental techniques. The crystallographic structure of the air-exposed sample was confirmed by powder x-ray diffraction at room temperature using Cu-K$_\alpha$ radiation. The crystals were exfoliated by standard scotch tape technique on Si/SiO$_2$ substrate, and the 2D-flakes were observed under an optical microscope. 2D layered structures of these single crystals were also confirmed using a scanning electron microscopy (JEOL JSM-6010LA), an atomic force microscopy (Asylum Research MFP-3D) and a transmission electron microscopy (JEOL JEM 2100F). Energy dispersive x-ray spectroscopy (EDS) study was carried out to know the elemental composition of the single crystals. Magnetic measurements of the bulk single crystals were performed using a SQUID magnetometer (MPMS XL, Quantum Design). Magneto-transport measurements were carried out using a physical property measurement system (PPMS II, Quantum Design).

## III. RESULTS AND DISCUSSION

**Figure 1(a)** depicts the well-known crystal structure of 2D layered $Cr_2Ge_2Te_6$ crystal where each unit cell consists of three $Cr_2Ge_2Te_6$ layers stacked along the *c* axis. Pristine $Cr_2Ge_2Te_6$ is a 2D layered lattice with a large interlayer separation of 6.9 Å which shows a hexagonal structure (*space group*: $R\bar{3}$, No. 148) with unit cell parameters of $a = b = 6.83$ Å, and $c = 20.56$ Å.[7] X-ray diffraction (XRD) patterns of the air-exposed single crystals are shown in **Fig. 1(b)** where only (00*l*) peaks are observed indicating that the crystal surface is normal to the *c* axis. A few degraded single crystals are further grinded into its powder form using a mortar-pestle and the obtained powder XRD pattern at room temperature is also displayed in **Fig. 1(b)**. The traces of a few peaks other than pristine $Cr_2Ge_2Te_6$ indicate the presence of an impurity phase along with the hexagonal phase of pristine $Cr_2Ge_2Te_6$ crystal. The estimated lattice parameters are found to be of $a = b = 6.89$ Å, and $c = 20.42$ Å. Hence, the lattice constant, *a*, of air-exposed

$Cr_2Ge_2Te_6$ enhances compared to pristine $Cr_2Ge_2Te_6$ system due to development of an impurity phase which in turn increases Cr-Cr distance. The increase in lattice parameter, *a*, can be analogous to a tensile strain applied to the system which can be assumed through the reduction of lattice constant, *c*, in the air-exposed crystal. A similar behavior in this type of system is observed when a strain is applied to a layered material[14]. Thus, the presence of an impurity phase can effectively modulate the magnetic properties of the studied system. The plate shapes of single crystals with an average size of 2-3 mm are shown in the **inset of Fig. 1(b)**. Scanning electron microscopy (SEM) micrograph of air-exposed single crystals is shown in **Fig. 1(c)** which reveals the clear vdW assembly of the layers. Atomic force microscopy image of an exfoliated flake on the $Si/SiO_2$ substrate is displayed in **Fig. 1(d)** where the line profile indicates different thickness of the flake and hence, the 2D vdW nature of the material. However, from the energy dispersive x-ray spectroscopy (EDS) analysis, the stoichiometric ratio of individual elements in the air-exposed crystal is found to be as Cr: Ge: Te ~ 1.33: 1: 3.54. Thus, the presence of excess Cr and Te in the degraded system leads us to find a possible stable binary phase of Cr and Te. From the crystallographic point of view, $Cr_2Te_3$ and $Cr_2Ge_2Te_6$ crystal have similar hexagonal structure and the conversion of $Cr_2Ge_2Te_6$ from $Cr_2Te_3$ can be realized if Ge atom can substitute selective Cr atom in $Cr_{III}$ site and one-half $Cr_{II}$ site of $Cr_2Te_3$ by breaking the Cr-Te ionic bonds.[20] Therefore, the possible impurity phase may be $Cr_2Te_3$ system. Recently, Yang et al.[20] proposed a selective ion exchange method of replacing magnetic $Cr^{3+}$ ions of non-vdW metallic $Cr_2Te_3$ by non-magnetic $Ge^{4+}$ until it completely becomes $Cr_2Ge_2Te_6$ nanosheets. Coexistence of both $Cr_2Te_3$ and $Cr_2Ge_2Te_6$ phases are well described for different intermediate phases of $Ge^{4+}$ doping. Interestingly, powder XRD pattern of the air-exposed $Cr_2Ge_2Te_6$ crystal is found to match well with the XRD pattern of an intermediate $Cr_2Te_3/Cr_2Ge_2Te_6$ phase as described in ref. 20.

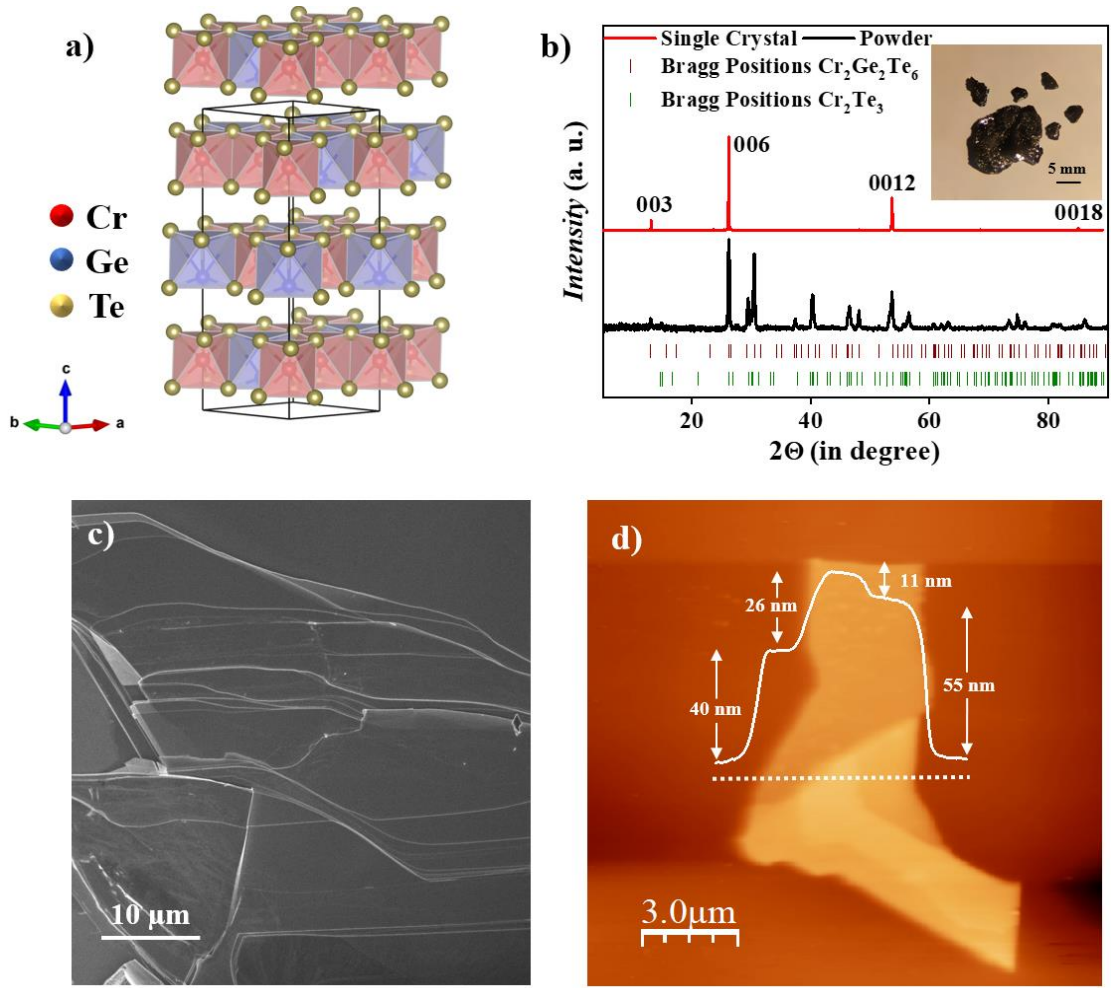

**Fig .1:** (a) Crystal structure of a pristine layered $Cr_2Ge_2Te_6$ crystals (b) Single crystal and powder XRD pattern of an air-exposed $Cr_2Ge_2Te_6$ crystal (inset: representative image of single crystals) (c) SEM micrograph of air-exposed $Cr_2Ge_2Te_6$ (d) AFM image of an exfoliated air-exposed $Cr_2Ge_2Te_6$ flake on $Si/SiO_2$ substrate.

In general, intrinsic 2D magnetic materials possess strong anisotropy between in-plane and out-of-plane physical properties and hence, the studied system can show strong magnetic anisotropy. From the temperature dependent *dc* magnetic susceptibility ($\chi$-*T*) curve of the pristine $Cr_2Ge_2Te_6$ as shown in **Fig. 2(a)**, it undergoes a continuous or second order paramagnetic (PM) to FM transition at $T_C \sim 67$ K which is in good agreement with previous studies[10–12]. Further, temperature dependence of magnetization (*M-T*) curve of the air-exposed $Cr_2Ge_2Te_6$ crystal is recorded at 100 Oe magnetic field under zero field cooled (ZFC) and field cooled (FC) condition with the applied field parallel to the *c* axis (*H//c*) and *ab* plane of the crystal (*H//ab*). χ-*T* curves of air-exposed $Cr_2Ge_2Te_6$ crystal are plotted in **Fig. 2(c)** and **Fig. 2(e)** for out-of-plane and in-plane direction of magnetic field respectively. For both cases with *H//c* and *H//ab*, χ shows a similar behavior where two magnetic ordering temperatures can be observed. In case of *H//c*, magnetic phase transitions are observed at about $T_{C1}$ ~ 69 K and $T_{C2}$ ~ 160 K which are estimated from the peak of dχ/d*T* vs. *T* plot as shown in the **inset of Fig. 2(c)** and these values are identical with those values obtained with *H//ab* as shown in **inset of Fig. 2(e)**. In addition, a short-range magnetic ordering is noticed near about 189 K and 184 K for the case of *H//c* and *H//ab* respectively. However, the change in susceptibility along the *c* axis and *ab* axis indicates that the system shows fair magnetic anisotropy which is strong enough compared to the pristine $Cr_2Ge_2Te_6$. The magnetic phase transition at about 69 K is related to the $T_C$ of the pristine $Cr_2Ge_2Te_6$ crystal whereas the following magnetic phase transition at 160 K is linked with the $T_C$ of ferromagnetic $Cr_2Te_3$ crystal[26]. Therefore, the enhancement of magnetic ordering towards higher temperature can be observed through multiple magnetic phase transitions which can be attributed to the increase in nearest neighbour Cr-Cr distance with the development of a residue phase of $Cr_2Te_3$ in the air-exposed $Cr_2Ge_2Te_6$ crystal.

From magnetic field dependent magnetization (*M-H*) plot at 3 K (shown in **Fig. 2(b)**) of pristine $Cr_2Ge_2Te_6$, it shows a typical soft FM behavior. Saturation magnetization ($M_S$) is estimated to be about 2.8 $\mu_B$/f.u. for the case of *H//c* whereas, it is ~2.9 $\mu_B$/f.u. for *H//ab*. These values are in good agreement to the expected value of 3 $\mu_B$ for $Cr^{3+}$ with three unpaired electrons. *M-H* curves for both *H//c* and *H//ab* show similar magnetization behavior though the saturation magnetic field ($H_S$) for *H//c* is smaller than *H//ab*. The small change in $H_S$ between these two cases indicates weak magnetic anisotropy. Similar behavior is reported in previous studies on $Cr_2Ge_2Te_6$[10,12]. For an anisotropic magnetic system, the uniaxial magnetocrystalline anisotropy constant ($K_u$) can quantify the magnetocrystalline energy and it can be determined using $\frac{2K_u}{M_S} = \mu_0 H_S$[16] where, $\mu_0$ is the vacuum permeability. The estimated $K_u$ about 9.53 J/kg at 3 K, which is in agreement with the previous studies[27], indicates that $Cr_2Ge_2Te_6$ crystal is a 2D FM with weak magnetic anisotropy.

   *M-H* curves of the air-exposed crystal at different temperatures are plotted in **Fig. 2(d)** and **Fig. 2(f)** for out-of-plane and in-plane direction of magnetic field respectively. From $\chi$-*T* and *M-H* plots, the air-exposed $Cr_2Ge_2Te_6$ crystal shows a typical FM like behavior. In case of *H//c*, at 3 K below $T < T_{C1}$ the material shows a strong FM with a large $M_S$ of about 44.3 emu/g which is larger compared to the value for pristine $Cr_2Ge_2Te_6$. In the intermediate temperature regime, $T_{C1} < T < T_{C2}$, FM ordering can be observed with a $M_S$ of about 36.2 emu/g at 100 K. The material exhibits PM like behavior with a linear magnetic field dependence of magnetization at 250 K which is well above of $T_{C2}$. Similar magnetic properties are observed for the case of *H//ab*. However, the magnetization in *M-H* curve changes in a two-steps process with the applied field. At low temperature of about 3 K, magnetization increases rapidly with field up to $H_1$ = 4.5 kOe and afterward, there is a change in slope of *M-H* curve where the magnetization increases slowly with magnetic field and shows a non-saturation tendency up to a maximum applied field of 50 kOe. The non-saturation magnetization even at application of high field indicates a frustrated FM interaction and a similar kind of behavior has been observed in $Cr_2Te_3$ system[26,28]. This kind of magnetization in *M-H* curve usually arises due to the coexistence of more than one magnetic phase and here, the possible coexisting phases are $Cr_2Ge_2Te_6$ with $Cr_2Te_3$. Moreover, magnetization increases sharply along *c* axis than along *ab* plane which indicates that the air-exposed $Cr_2Ge_2Te_6$ system shows strong magnetic anisotropy. The $K_u$ is calculated considering $\mu_0 H_S \sim 5$ T though it is larger than 5 T in real case. Thus, the underestimated value of $K_u$ for air-exposed $Cr_2Ge_2Te_6$ is found to be 110.7 J/kg which is significantly stronger compared to the estimated value for pristine $Cr_2Ge_2Te_6$. The enhanced magnetic anisotropy in the air-exposed $Cr_2Ge_2Te_6$ can be attributed to the presence of $Cr_2Te_3$ phase with strong magnetic anisotropy.

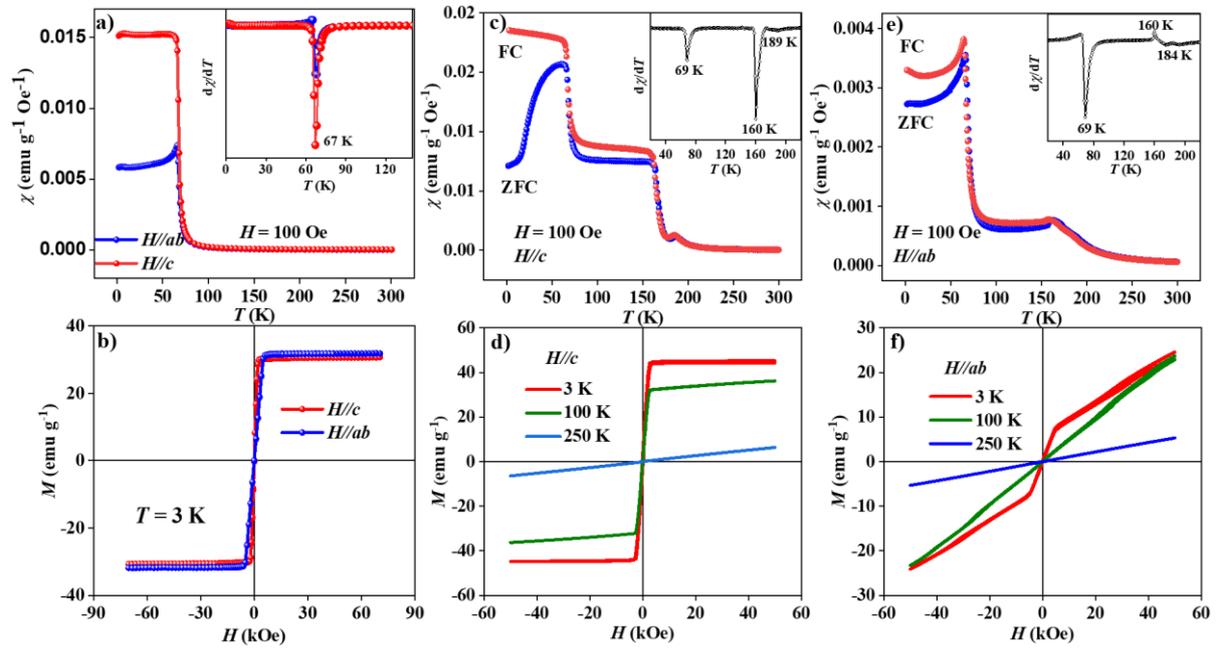

**Fig. 2: Magnetic properties of pristine $Cr_2Ge_2Te_6$.** (a) χ vs. T plot in presence of 100 Oe magnetic field during ZFC condition with *H//c* and *H//ab* of the pristine crystal (**inset:** dχ/d*T* vs. *T* plot to deduce the magnetic phase transition temperature) (b) M-H curve at 3 K with *H//c* and *H//ab* conditions. **Magnetic properties of air-exposed $Cr_2Ge_2Te_6$.** (c) χ-T plot in presence of 100 Oe magnetic field during both ZFC and FC condition with *H//c* of the air-exposed crystal (**inset:** dχ/d*T* vs. *T* plot) (d) *M-H* at different temperatures with *H//c* (e) χ-*T* plot with *H//ab* of the air-exposed crystal during ZFC and FC condition (**inset:** dχ/d*T* vs. *T* plot) (f) *M-H* plot at various temperatures with *H//c* condition.

From the χ-*T* plot, the presence of large bifurcation and the non-saturation magnetization behavior in *M-H* plot with *H//ab* lead us to find the possibility of existence of spin-glass phase in the air-exposed $Cr_2Ge_2Te_6$. In general, the shifting of phase transition temperatures with frequencies in temperature dependent ac magnetic susceptibility, long time relaxation of isothermal remanent magnetization and memory effects are the fingerprints of spin-glass state at low temperature. **Figure 3** represents the real part of ac magnetic susceptibility (χ′) as a function of temperature of the air-exposed $Cr_2Ge_2Te_6$ crystal recorded at various frequencies ranging from 77 Hz to 1277 Hz. From **Fig. 3**, both, $T_{C1}$ and $T_{C2}$ are insensitive at different frequencies which implies that spin-glass state is not present in the system. The estimated values of $T_{C1}$ and $T_{C2}$ from $\chi'$-$T$ are in good agreement with the deduces values from d$\chi$/d$T$ vs. $T$ plot as displayed in **Fig. 2(c)** and **2(e)**. Moreover, isothermal remanent magnetization ($M_{IRM}$) of the sample is plotted as a function of time in **inset of Fig. 3**. The protocol to measure $M_{IRM}$ was the following: 1) starting at high temperature (T > 200K), the sample was cooled down at zero magnetic field (ZFC) to the specified temperature (2 K); 2) a magnetic field of 1 T was applied for half an hour; 3) the magnetic field was brought back to zero, and $M_{IRM}$ was recorded as function of time (t). t = 0 was defined as the instant when the field reached zero. $M_{IRM}(t)$ decays exponentially with time and is found to follow the relationship[29]

$$M_{IRM}(t) = M_0(t) - S(T)\ln\left(1 + \frac{t}{t_0}\right) \quad (1)$$

where, $M_0(t)$ and $S(T)$ are the initial magnetization at $t$ = 0 and magnetic viscosity respectively. The term $t_0$ differs on the measuring conditions and has limited physical relevance. This dependence is expected for thermally activated magnetization reversals over activation barriers.[30] $M_{IRM}$ undergoes a slow relaxation with time where the value of $M_{IRM}$ reduces by about 2.8% from its initial value after 100 minutes. Moreover, the estimated value of $S(T)$ about 0.02011 emu/g from the fitting is very similar to the obtained value of $S(T)$ in case of already explored Y-substituted $Gd_2PdSi_3$ alloys which is free from the spin-glass state at low temperature region.[29] Besides, relaxation behavior is a typical feature of FM materials, and the decay behavior in $M_{IRM}$ can be due to the short-range magnetic correlation between two phases at below 5 K.

To get phenomenological understanding of the coexisting long-range magnetic orders in a single system, Ginzburg-Landau type free energy expression can be written with two independent order parameters ($m_1$ and $m_2$) corresponding to the mean-field transition temperatures ($T_{C1}$ and $T_{C2}$, $T_{C2}$ > $T_{C1}$) with a coupling term c,

$$F = V\left[\frac{a_1}{2}m_1^2 + \frac{b_1}{4}m_1^4 + \frac{a_2}{2}m_2^2 + \frac{b_2}{2}m_2^4 + \frac{c}{2}m_1^2m_2^2 - (m_1 + m_2)h\right] \quad (2)$$

The coefficients $a_{1,2}$ are temperature dependent $a_{1,2} = \alpha_{1,2}(T - T_{C1,C2})$ with $T_{C2} > T_{C1}$. The remaining coefficients are temperature independent. The minimization of the free energy at zero external field gives the first transition at $T_{C2}$ unaffected by the coupling because it is higher than $T_{C1}$. The magnetization $m_1$ becomes nonzero at $T_{C1}' = T_{C1}(1 - \beta T_{C2}/T_{C1})/(1 - T_{C2}/T_{C1})$ where $\beta = c\alpha_1/b\alpha_2$. This transition temperature can be lower or higher than the assumed transition temperature $T_{C1}$ depending on the sign of the coupling 'c'. However, it must be borne in mind that transition temperatures can only be measured, and we simply set the actual transition temperature as $T_C = T_{C1}'$. Interestingly, the sign of 'c' determines whether the extra addition will be large or small. Besides, the presence of short-range interaction, originating from magnetic exchange interaction within phase boundaries, is responsible for coexistence of two attributed phases.

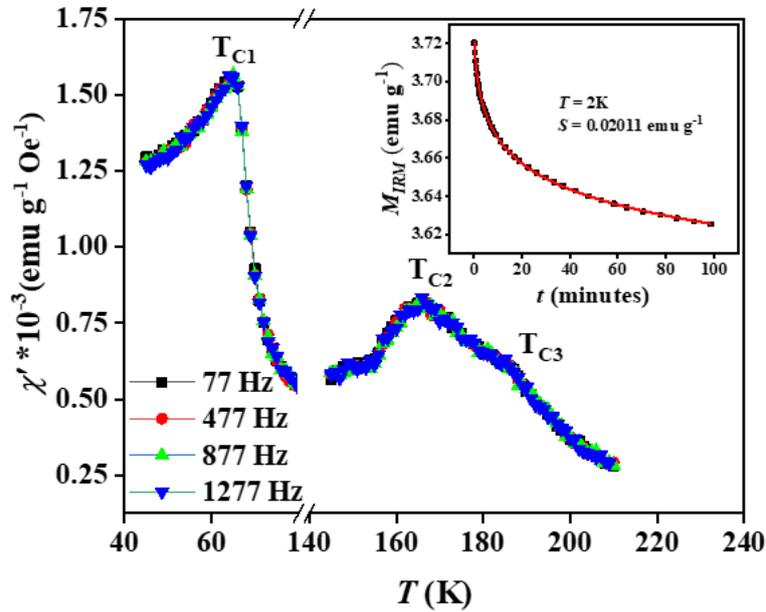

**Fig .3:** Temperature dependence of real part of ac susceptibility under different applied frequencies. [**inset:** isothermal remanent magnetization vs. time t at 2 K. The solid line represents the fitting using the expression (1)].

Further, the temperature dependence of zero-field specific heat capacity ($C_p$) data is recorded for both pristine and air-exposed $Cr_2Ge_2Te_6$ crystal (shown in supporting information, **Fig. S1**). $Cp$ shows a clear corelation to the measured magnetization data as displayed in **Fig. 2(c) and 2(e)** for air exposed $Cr_2Ge_2Te_6$ and in **Fig. 2(a)** for pristine crystal. The pristine $Cr_2Ge_2Te_6$ shows a Λ-shape peak near $T_C$ value which is a signature of second-order PM to FM transition. In case of air-exposed $Cr_2Ge_2Te_6$, $C_p$ shows three different peaks near $T_{C1}$, $T_{C2}$ and at $T_{C3}$ ~ 181 K and the values are in good agreement with the observed magnetic phase transition temperatures from $\chi$-$T$ plot.

Due to the presence of magnetocrystalline anisotropy and the coexistence of two phases in the air-exposed $Cr_2Ge_2Te_6$, it would be interesting to probe magnetotransport. Therefore, we measured isothermal resistivity with varying magnetic field. **Figure 4** shows magnetoresistance (MR) of the sample, measured at 5 K with the magnetic field along the *c* axis and *ab* plane of the sample following the equation, $MR = \frac{(R(H)-R(0))}{R(0)} \times 100\ \%$. $R(H)$ and $R(0)$ represent the measured electrical resistances at finite and zero magnetic field respectively. To measure the electrical resistance, the sample is cooled down to the targeted temperature from room temperature under ZFC condition. From magnetic field direction dependence of MR, air-exposed $Cr_2Ge_2Te_6$ crystal shows anisotropic MR, and the effective MR shows a combine feature of MR behavior of two individual $Cr_2Te_3$ and $Cr_2Ge_2Te_6$ phases which is discussed in the following section. The anisotropic MR in the system arises from the anisotropy of spin-orbit interaction of the material.[11,31]

From **Fig. 4(a)**, with the magnetic field applied along the *c* axis, magnetoresistance at 5 K shows a positive slope with increase in magnetic field within +/- 0.45 T which seems like parabolic dependence on magnetic field. The low-field MR data can be fitted with second-order polynomial expression, $MR\ (H) = aH^2 + bH + c$ which is a simplified form of the model proposed for spin disorder scattering[32,33] and the parameters *a*, *b*, *c* roughly contains the terms such as, Fermi wave number ($k_F$) the elementary charge (e) carrier effective mass (m), Planck's constant (h), density ($N_v$) and spin (S) of chromium ions. At higher fields, the curve shows a linear and negative slope. Further, MR is measured at high temperature of about 25 K and 40 K. MR is positive for all the temperatures and its magnitude is less than 1% at 5 K which enhances with increasing temperature. This type of magnetoresistance behaviour is attributed to the short-range interaction between different magnetic domains present in the crystal[34,35]. At higher temperatures (shown in supporting information, **Fig. S2**), MR shows similar behavior but the magnetic field where the slope of MR changes from positive to negative increases to a higher value (~ 0.84 T for 25 K) as the domain wall ordering is restricted by the thermal fluctuation. At 40 K, MR shows a positive slope in the entire field range of 0-2 T and thus, negative slope of MR is expected at higher field beyond 2 T as the applied magnetic field is not enough to counteract thermal fluctuation resisting the domain alignment. However, for pure $Cr_2Te_3$ system, MR at low temperature consists of two parts[36]: in the high field region, it shows linear and reversible negative MR whereas in the low field region, MR shows two sharp maxima during positive and negative field sweeping which are correspond to the coercive field of $Cr_2Te_3$. Thus, the negative slope in MR at higher field in case of air-exposed $Cr_2Ge_2Te_6$ arises due to the presence of a residue of $Cr_2Te_3$ phase.

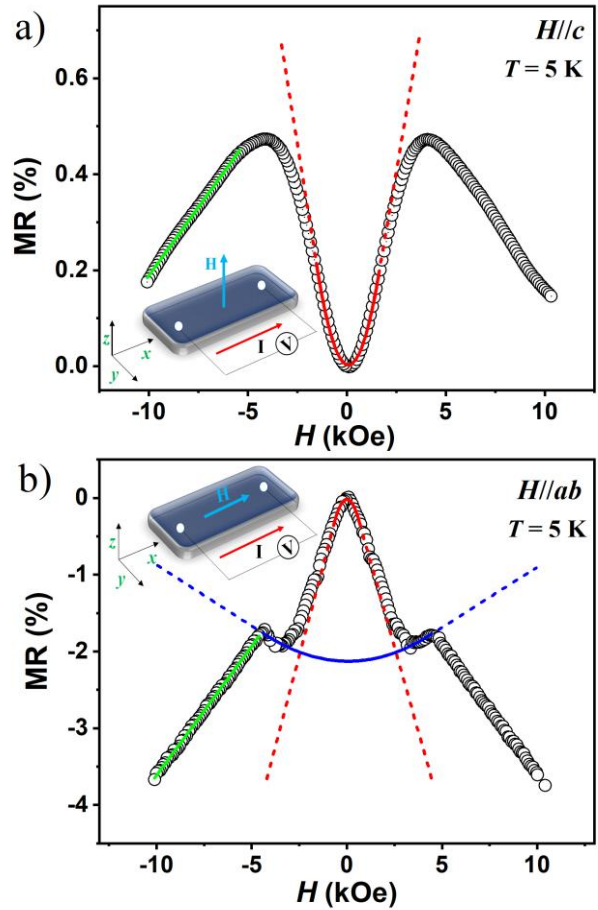

**Fig. 4:** Magnetic field dependent of magnetoresistance at 5 K (a) with the magnetic field is applied along *c* axis (b) with the magnetic field is applied parallel to *ab* plane. Black circular points are experimental data. Solid lines (red and blue colour) are the fits in the low field region following the expression, $MR(H) = aH^2 + bH + c$. Dash lines are linear extrapolation from the fits. Solid green lines are the linear fits in the high field region following $MR(H) = bH + c$.

With the magnetic field parallel to *ab* axis, the crystal shows a negative MR with increasing the field at 5 K as shown in **Fig. 4(b)**. In the low field region, MR data are found to be fitted with the above-mentioned second order polynomial expression. At ± 0.45 T, where the slope of magnetization changes in *M-H* plot for *H//ab*, a change in slope of MR from negative to positive is observed which vanishes at higher temperature (shown in supporting information, **Fig. S2(b)**). This change in slope of MR with the magnetic field can also be fitted with the above expression. Afterwards, at higher fields it shows a linear and negative slope which is due to the suppression of spin-flip scattering at high magnetic fields. Maximum negative value of MR at 40 K is found to be about – 50 % for a field change of 2 T. However, in case of the pristine $Cr_2Te_3$ system with strong magnetic anisotropy, MR shows firstly a positive trend as long as magnetic field increases to the saturation field and subsequently, at higher field MR shows a negative and linear slope.[36] Therefore, the present air-exposed $Cr_2Ge_2Te_6$ system shows an effective MR which associates two different features of two above-mentioned coexisting phases. Additionally, Angle-resolved photoemission spectroscopy using a micro-focused beam spot (micro-ARPES) or X-ray absorption spectroscopy (XAS) study will be helpful for understanding the local electronic structure of the hybrid system and the corelation between crystal phase transformation and associated magnetic interaction from the fundamental aspects.

## IV. CONCLUSION

In summary, ferromagnetic ordering of $Cr_2Ge_2Te_6$ phase is manipulated by spontaneously occurring a high Curie temperature magnetic phase through developing a residue phase of non-vdW ferromagnetic $Cr_2Te_3$ in an air-exposed $Cr_2Ge_2Te_6$ crystal. The air-exposed $Cr_2Ge_2Te_6$ crystal is found to exhibit ferromagnetic ordering below 160 K which is largely enhanced from the pristine $Cr_2Ge_2Te_6$ phase. Though the developed phase is non-vdW in its pristine form, the air-exposed $Cr_2Ge_2Te_6$ system is nicely exfoliable showing vdW nature. Moreover, magnetic anisotropy is considerably enhanced in air-exposed $Cr_2Ge_2Te_6$ system compared to the pristine bulk with weak magnetic anisotropy. Therefore, the enhancement of ferromagnetic Curie temperature with strong magnetic anisotropy in $Cr_2Ge_2Te_6$ increases its potential for the application in low dimensional nanoelectronics and spintronics devices.

**ACKNOWLEDGEMENT**

Authors would like to thank Prof. Saurav Giri for the magnetotransport measurements. BD and MP are grateful to IACS for the fellowship. TK and SM acknowledge DST-INSPIRE for their fellowship. RP is grateful to CSIR for the fellowship. SD acknowledges the financial support from DST-SERB grant No. CRG/2021/004334. SD also acknowledges support from the Technical Research Centre (TRC), IACS, Kolkata.

SD and BD conceived the project and designed the experiments. RM prepared the samples and performed their initial characterization. BD carried out all the measurements with time-elapsed crystal in consultation with SG, TK, RP, SM and MP. BD performed the magnetotransport measurements with TKD. Confirmatory experiments regarding magnetization of the aged crystal were recorded by SS, PAS and MM. BD and SG analyzed the XRD data with RAS. JKB developed the analytical model with contributions from BD, and S.D. All authors discussed the results and actively commented on the manuscript written by BD and SG with the input from SD.

# Supporting Information

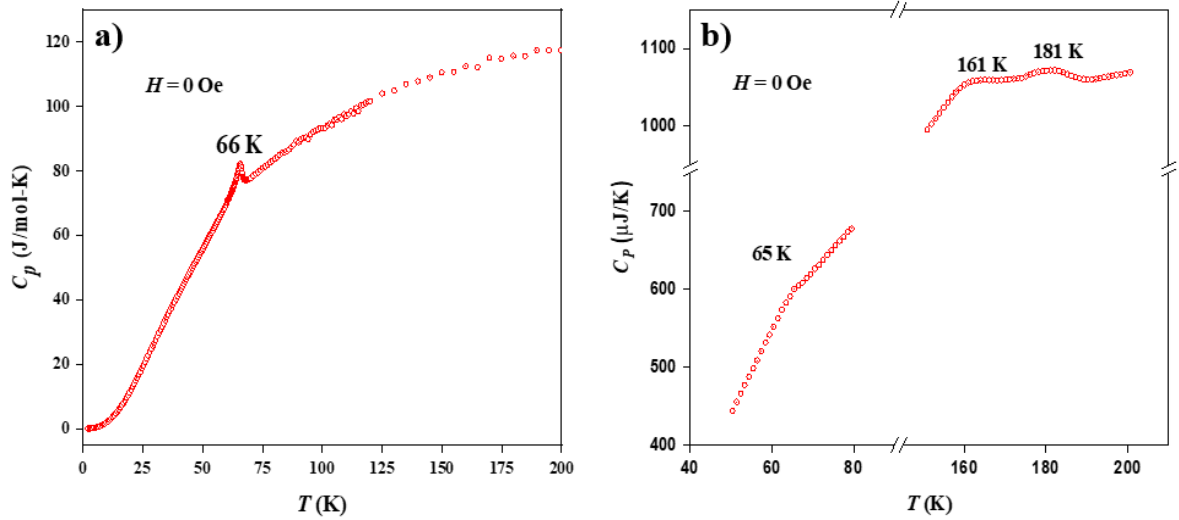

**Fig. S1:** (a) Specific heat capacity ($C_p$) of pristine $Cr_2Ge_2Te_6$ as a function of temperature recorded in zero field. (b) Temperature dependence of zero field $C_p$ for air-exposed $Cr_2Ge_2Te_6$ crystal.

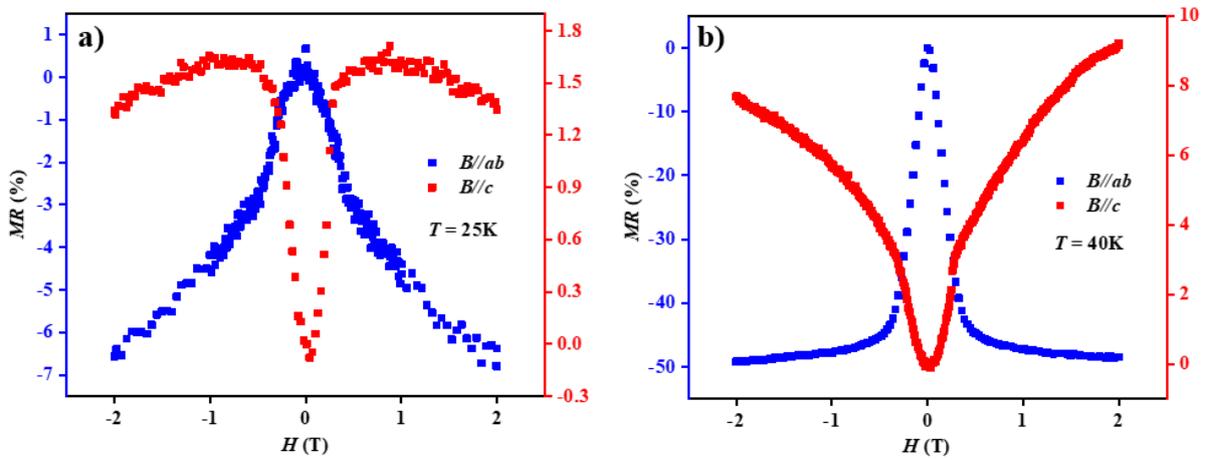

**Fig. S2:** MR as a function of applied magnetic field with *H*//*c* and *H*//*ab* at (a) 25 K and (b) 40 K.